\def\grtsim{{~\raise.15em\hbox{$>$}\kern-.85em \lower.35em\hbox{$\sim$}~}}
\def\lesssim{{~\raise.15em\hbox{$<$}\kern-.85em \lower.35em\hbox{$\sim$}~}}
\def\tighten{}
\def\preprint#1{\rightline{#1}}
\def\draft{}
\def\mnewpage{}          % no separate first page
\def\gtrsim{{~\raise.15em\hbox{$>$}\kern-.85em\lower.35em\hbox{$\sim$}~}}
\documentstyle[sprocl]{article}

\newcommand{\beq}{\begin{equation}}
\newcommand{\eeq}{\end{equation}}
\newcommand{\beqa}{\begin{eqnarray}}
\newcommand{\eeqa}{\end{eqnarray}}
\newcommand{\klpnn}{\mbox{$K_L \to \pi^0 \nu \bar \nu$}}
\newcommand{\klpnnij}{\mbox{$K_L \to \pi^0 \nu_i \bar \nu_j$}}

\newcommand{\kspnn}{\mbox{$K_S \to \pi^0 \nu \bar \nu$}}

\newcommand{\kppnn}{\mbox{$K^+ \to \pi^+ \nu \bar \nu$}}

\newcommand{\kzpnn}{\mbox{$K^0 \to \pi^0 \nu \bar \nu$}}
\newcommand{\kpnn}{\mbox{$K \to \pi \nu \bar \nu$}}

\def\O{{\cal O}}
\def\epsK{\varepsilon}
\def\BR{\mbox{{\rm BR}}}
\def\SM{Standard Model}
\def\NP{New Physics}
\def\Re{\mbox{{\rm Re}}}
\def\Im{\mbox{{\rm Im}}}
\def\ris{r_{\rm is}}

\def\npb#1{Nucl.\ Phys.\ {\bf B #1}}
\def\plb#1{Phys.\ Lett.\ {\bf B #1}}
\def\prd#1{Phys.\ Rev.\ {\bf D #1}}
\def\prl#1{Phys.\ Rev.\ Lett. {\bf #1}}

\begin{document}

\draft

{\tighten
\preprint{
\vbox{
      \hbox{SLAC-PUB-7443}
      \hbox{hep-ph/9704208}}}

\bigskip
%\bigskip

\renewcommand{\thefootnote}{\fnsymbol{footnote}}

\title{The \klpnn\ Decay Beyond the Standard Model}
%\footnotetext{Research supported
%by the Department of Energy under contract DE-AC03-76SF00515}
\author{Yuval Grossman}
\footnotetext{Talk given at the first symposium of FCNC, February 19-21,
     1997, Santa Monica, CA.}
\address{ \vbox{\vskip 0.truecm}
%$^a$
Stanford Linear Accelerator Center \\
        Stanford University, Stanford, CA 94309 }
%\vbox{\vskip 0.truecm}
%  $^b$Department of Particle Physics \\
%  Weizmann Institute of Science, Rehovot 76100, Israel}

\maketitle

\begin{abstract}%
The \klpnn\ decay is analyzed in a model independent way.
When lepton flavor is conserved, this decay mode is a manifestation of CP
violating interference between mixing and decay. Consequently, a
theoretically clean relation between the measured rate and electroweak
parameters holds in any given model.

\end{abstract}

} % end tighten

\mnewpage

%%%%%%%%%%%%%%%%%%%%%%%%%%%%%%%%%
%%%%%%%%%%%%%   I   %%%%%%%%%%%%%
%%%%%%%%%%%%%%%%%%%%%%%%%%%%%%%%%
%\section{Introduction}

\klpnn\ is unique among $K$ decays in several aspects: (a) It is
theoretically very clean; (b) it is purely CP violating \cite{Litt,BuBu};
and (c) it can be measured in the near future \cite{exper}\
even if the rate is as small as the \SM\ prediction.
In the \SM\ a measurement of $\Gamma(\klpnn)$ provides a clean
determination of the Wolfenstein CP violating parameter $\eta$ or,
equivalently, of the Jarlskog measure of CP violation $J$ and, together
with a measurement of $\Gamma(\kppnn)$,
of the angle $\beta$ of the unitarity triangle \cite{BuBu}.

Here we explain what can be learned from the \kpnn\ decay in a model
independent way \cite{klnn}. We define
\beq \label{deflam}
\lambda \equiv {q \over p} {\bar A \over A},
\eeq
where $p$ and $q$ are the components of interaction
eigenstates in mass eigenstates,
$|K_{L,S}\rangle=p |K^0 \rangle \mp q |\bar K^0 \rangle$, and
$A(\bar A$) is the $K^0(\bar K^0)\to \pi^0\nu\bar\nu$ decay amplitude.
Then, the ratio between the $K_L$ and $K_S$ decay rates is \cite{klnn}
\beq \label{ratrat}
{\Gamma(\klpnn) \over \Gamma(\kspnn)} =
{1 + |\lambda|^2 - 2\Re\lambda\over 1 + |\lambda|^2 + 2\Re\lambda}.
\eeq
In general, a three body final state does not have a definite CP parity.
However, if the light neutrinos are purely left-handed, and if lepton
flavor is conserved, the final state is CP even (to an excellent
approximation) \cite{klnn}. If lepton flavor is violated, the final state
in \klpnn\ is not necessarily a CP eigenstate; specifically, $\klpnnij$
with $i \neq j$ is allowed. Here, we concentrate on the case where the
above two conditions are satisfied, so that the final state is purely CP
even.

The contributions to the \klpnn\ decay from CP violation in mixing
($|q/p|\neq1$) and from CP violation in decay ($|\bar A/A|\neq1$) are
negligibly small. The deviation of $|q/p|$ from unity is experimentally
measured (by the CP asymmetry in $K_L\to\pi\ell\nu$)
and is $\O(10^{-3})$. The deviation of $|\bar A/A|$ from unity is
expected to be even smaller \cite{klnn}. Therefore, $|\lambda|=1 +
\O(10^{-3})$, and the leading CP violating effect is $\Im\lambda\neq0$,
namely interference between mixing and decay. This puts the ratio of
decay rates (\ref{ratrat}) in the same class as CP asymmetries in various
$B$ decays to final CP eigenstates, e.g. $B\to\psi K_S$, where a very
clean theoretical analysis is possible \cite{Breviw}.

As a result of this cleanliness, the CP violating phase can be extracted
almost without any hadronic uncertainty,
even if this phase comes from
\NP. Defining $\theta$ to be the relative phase between the
$K-\bar K$ mixing amplitude and the $s\to d\nu\bar\nu$ decay amplitude,
namely $\lambda=e^{2i\theta}$, we get from eq. (\ref{ratrat})
\beq \label{genrat}
{\Gamma(\klpnn) \over \Gamma(\kspnn)} =
{1 - \cos 2 \theta \over 1 + \cos 2 \theta} = \tan^2 \theta.
\eeq
In reality, however, it will be impossible to
measure $\Gamma(\kspnn)$. We can use the isospin relation,
$A(\kzpnn)/A(\kppnn)=1/\sqrt{2}$,
to replace the denominator by the charged kaon decay mode:
\beq \label{ratis}
a_{CP} \equiv \ris {\Gamma(\klpnn) \over \Gamma(\kppnn)} =
{1 - \cos 2 \theta \over 2} = \sin^2 \theta,
\eeq
where $\ris = 0.954$ is the isospin breaking factor \cite{MaPa}.
The ratio (\ref{ratis}) may be experimentally measurable as the relevant
branching ratios are $\O(10^{-10})$ in the \SM\ \cite{BuBu}
and even larger in some of its extensions.

%The important consequence of 
Eq. (\ref{ratis}) implies that a measurement of
$a_{CP}$ will allow us to determine the CP violating phase $\theta$
without any information about the magnitude of the decay amplitudes.
%The ratio (\ref{ratis}) is most useful if both \klpnn\ and \kppnn\ are
%dominated by the same combination of mixing angles. The phase of this
%combination is then directly identified with $\theta$, and we need not
%know any other of the new parameters.
Also, using
$\sin^2 \theta \le 1$ and $\tau_{K_L}/\tau_{K^+}=4.17$, we get the 
model independent bound
\beq
\BR(\klpnn) < 1.1 \times 10^{-8} 
\left({\BR(\kppnn) \over 2.4 \times 10^{-9}}\right).
\eeq
This bound is much stronger than the direct experimental upper bound
\cite{Weaver} $\BR(\klpnn) < 5.8 \times 10^{-5}$.

\NP\ can modify both the mixing and the decay amplitudes.
%The contribution to the mixing can be of the same order as the \SM\ one.
$\epsK=\O(10^{-3})$ implies that any new contribution
to the mixing amplitude carries almost the same phase as the \SM\ one.
On the other hand, the upper bound \cite{Adler}
$\BR(\kppnn) < 2.4 \times 10^{-9}$,
which is much larger than the \SM\ prediction \cite{BuBu}, allows
\NP\ to dominate the decay amplitude (with an arbitrary phase). We
conclude that a significant modification of $a_{CP}$ can only come from
\NP\ in the decay amplitude. For example, in models with extra
quarks, the decay amplitudes can be dominated by tree level $Z$-mediated
diagrams \cite{klnn}.

In superweak models, all CP violating effects appear in the mixing
amplitudes. Then, CP violation in \klpnn\ should be similar in magnitude
to that in $K_L\to\pi\pi$. In models of approximate CP symmetry,
all CP violating effects are small. Both scenarios predict then
$a_{CP}=\O(10^{-3})$, in contrast to the \SM\ prediction, $a_{CP}=\O(1)$.
In other words, a measurement of $a_{CP}\gg10^{-3}$ (and, 
in particular,
$\BR(\klpnn)\gtrsim\O(10^{-11}))$ will exclude these two scenarios
of \NP\ in CP violation.

In the \SM\ there are two clean ways to determine the unitarity triangle:
(1) CP asymmetries in $B^0$ decays \cite{Breviw}; and (2) the combination
of $\BR(\klpnn)$ and $\BR(\kppnn)$ \cite{BuBu}. In general, \NP\
will affect both determinations. Moreover, it is very unlikely that
the modification of the two methods will be the same. Consequently,
a comparison between these two clean determinations will be a very
powerful tool to probe CP violation beyond the \SM. Because of the very
small theoretical uncertainties in both methods even a small new physics
effect can be detected. In practice, we will be limited only by the
experimental sensitivity.

In conclusion: a measurement of $\BR(\klpnn)$ is guaranteed to provide us
with valuable information. It will either give a new clean measurement of
CP violation or indicate lepton flavor violation.

%\mnewpage

{\bf Acknowledgments.}
%\acknowledgements
I thank Yossi Nir for collaboration on this work.
Y.G. is supported by the Department of Energy under contract
DE-AC03-76SF00515. 
%Y.N. is supported in part by the United States --
%Israel Binational Science Foundation (BSF), by the Israel Science
%Foundation, and by the Minerva Foundation (Munich).

{\tighten
%\begin{references}

%\end{references}
}


\begin{thebibliography}{99}

\bibitem{Litt}
{L.S. Littenberg, \prd{39} (1989) 3322.}

\bibitem{BuBu}
{G. Buchalla and A.J. Buras, \npb{400} (1993) 225; \prd{54} (1996) 6782;
A.J. Buras, \plb{333} (1994) 476; G. Buchalla, these proceedings.}


\bibitem{exper}
{T. Inagaki, these proceedings; D. Bryman, these proceedings;
K. Arisaka, these proceedings.}

\bibitem{klnn}
Y. Grossman, and Y. Nir, hep-ph/9701313, to appear in \plb{}.


\bibitem{Breviw}
%For a review 
See, e.g.
%Y. Nir, Lectures presented in the 20th SLAC Summer Institute,
%SLAC-PUB-5874 (1992);
Y. Nir and H.R. Quinn, Ann. Rev. Nucl. Part. Sci. {\bf 42} (1992) 211.

\bibitem{MaPa}
{W. Marciano and Z. Parsa, \prd{53} (1996) 1.}

\bibitem{Weaver}
{M. Weaver {\it et al.}, E799 Collaboration, \prl{72} (1994) 3758.}

\bibitem{Adler}
{S. Adler {\it et al.}, BNL 787 Collaboration, \prl{76} (1996) 1421.}

\end{thebibliography}
\end{document}